\documentclass[conference,a4paper]{IEEEtran}
\usepackage{amsfonts, amssymb, amsopn, amsthm, amsmath}
\usepackage[dvips]{graphicx}

\newtheorem{thm}{Theorem}

\newtheorem{claim}[thm]{Claim}
\providecommand{\abs}[1]{\lvert#1\rvert}

\newcommand{\esp}{\ensuremath{\mathbb{E}}}

\def\l2lim{\,{\buildrel L_2 \over \rightarrow}\,}

\def \EE{\mathbb E}

\title{Spatial Degrees of Freedom of MIMO Systems\\ in Line-of-Sight Environment}
\author{\IEEEauthorblockN{Marc Desgroseilliers, Olivier L\'ev\^eque, Emmanuel Preissmann}
\IEEEauthorblockA{School of Computer and Communication Sciences\\
Swiss Federal Institute of Technology - Lausanne, 1015 Lausanne, Switzerland\\
Emails: \{marc.desgroseilliers, olivier.leveque\}@epfl.ch, emmanuel.preissmann@gmail.com}}
\begin{document}

\maketitle

\begin{abstract} THIS PAPER IS ELIGIBLE FOR THE BEST STUDENT PAPER AWARD.
While the efficiency of MIMO transmissions in a rich scattering environment has been demonstrated, less is known about the situation where  the fading matrix coefficients come from a line-of-sight model. In this paper, we study in detail how this line-of-sight assumption affects the performance of distributed MIMO transmissions between far away clusters of nodes in a wireless network. Our analysis pertains to the study  of a new class of random matrices.
\end{abstract}

\section{Introduction}
The aim of the present paper is to study the number of spatial degrees of freedom of multiple-input multiple-output (MIMO) transmissions in a wireless network with homogeneously distributed nodes, under the following classical line-of-sight propagation model between node $k$ and node $j$ in the network:
\begin{equation} \label{eq:los_model}
h_{jk} = \frac{e^{2 \pi i r_{jk} / \lambda}}{r_{jk}}.
\end{equation}
In the above equation, $\lambda$ is the carrier wavelength and $r_{jk}$ is the internode distance. From a mathematical point of view, these matrices are interesting objects, as they are halfway between purely random matrices with i.i.d.~entries and fully deterministic matrices. Indeed, the internode distances $r_{jk}$ are random due to the random node positions, but there is a clear correlation between the matrix entries. 

Let us recall that the degrees of freedom of a MIMO transmission are defined as the number of independent streams of information that can be conveyed simultaneously and reliably over the channel at high SNR. Under the assumption of a channel fading matrix $H$ with i.i.d.~entries, this number of degrees of freedom is directly proportional to the number of antennas used for transmission and reception \cite{T99}.

The performance of MIMO systems in line-of-sight environment has been analyzed by various authors (see e.g.~\cite{outdoorMIMO,sayeed}) in the literature. Our intention here is to study this performance in the context of wireless networks, where large clusters of nodes are used as {\em virtual multiple antenna arrays}. In this case, MIMO transmissions may not benefit from all possible  degrees of freedom; it was indeed observed in \cite{FMM09} that under the above propagation model \eqref{eq:los_model}, MIMO transmissions suffer from a {\em spatial limitation}; if $A$ denotes the network area, $n$ the number of nodes in the network (assumed to be uniformly distributed) and $\lambda$ the carrier wavelength, then the number of spatial degrees of freedom of {\em any} MIMO transmission in the network cannot exceed\footnote{See \cite{FMM09} for a precise statement.}
\begin{equation} \label{eq:ub_dof}
\min \left(n, \sqrt{A}/\lambda \right).
\end{equation}
In case the network area $A$ remains reasonably large, this does not prevent the possibility of transmissions with full degrees of freedom in the network. Yet, transmissions between clusters of nodes confined to smaller areas and moreover separated by long distances may suffer from even more spatial limitations.

\begin{center}
\scalebox{1}{\begin{picture}(0,0)%
\includegraphics{squares.pstex}%
\end{picture}%
\setlength{\unitlength}{2763sp}%
\begingroup\makeatletter\ifx\SetFigFont\undefined%
\gdef\SetFigFont#1#2#3#4#5{%
  \reset@font\fontsize{#1}{#2pt}%
  \fontfamily{#3}\fontseries{#4}\fontshape{#5}%
  \selectfont}%
\fi\endgroup%
\begin{picture}(4902,2199)(1936,-2473)
\put(2476,-2386){\makebox(0,0)[lb]{\smash{{\SetFigFont{11}{13.2}{\familydefault}{\mddefault}{\updefault}{$\sqrt{A}$}
}}}}
\put(5776,-511){\makebox(0,0)[lb]{\smash{{\SetFigFont{11}{13.2}{\familydefault}{\mddefault}{\updefault}{$D_R$}
}}}}
\put(4276,-1036){\makebox(0,0)[lb]{\smash{{\SetFigFont{11}{13.2}{\familydefault}{\mddefault}{\updefault}{$d$}
}}}}
\put(1951,-886){\makebox(0,0)[lb]{\smash{{\SetFigFont{11}{13.2}{\familydefault}{\mddefault}{\updefault}{$D_T$}
}}}}
\put(5701,-2386){\makebox(0,0)[lb]{\smash{{\SetFigFont{11}{13.2}{\familydefault}{\mddefault}{\updefault}{$\sqrt{A}$}
}}}}
\end{picture}
}\\
Fig.~1.~Two square clusters of area $A$ separated by distance $d$.
\end{center}

In \cite{LC12, OLT13}, it was shown independently that for two clusters of area $A$ separated by distance $d$, as illustrated on Fig.~1, at least the following spatial degrees of freedom could be achieved\footnote{See Section \ref{sec:spat}, Theorem \ref{thm:lb} for a precise statement.}:
\begin{equation} \label{eq:lb_dof}
\begin{cases}
\min \left(n, \sqrt{A}/\lambda \right), & \quad\textrm{when}\quad 1 \le d \le \sqrt{A},\\
\min \Big( n, A/\lambda d \Big), & \quad\textrm{when}\quad \sqrt{A} \le d \le A/\lambda.
\end{cases}
\end{equation}
The situation is summarized on the graph below:

\begin{center}
\scalebox{0.5}{\begin{picture}(0,0)%
\includegraphics{scaling.pstex}%
\end{picture}%
\setlength{\unitlength}{3947sp}%
\begingroup\makeatletter\ifx\SetFigFont\undefined%
\gdef\SetFigFont#1#2#3#4#5{%
  \reset@font\fontsize{#1}{#2pt}%
  \fontfamily{#3}\fontseries{#4}\fontshape{#5}%
  \selectfont}%
\fi\endgroup%
\begin{picture}(7443,4051)(211,-4052)
\put(226,-1036){\makebox(0,0)[lb]{\smash{{\SetFigFont{17}{20.4}{\familydefault}{\mddefault}{\updefault}{$\sqrt{A}/\lambda$}%
}}}}
\put(2176,-3961){\makebox(0,0)[lb]{\smash{{\SetFigFont{17}{20.4}{\familydefault}{\mddefault}{\updefault}{$\sqrt{A}$}%
}}}}
\put(5851,-3886){\makebox(0,0)[lb]{\smash{{\SetFigFont{17}{20.4}{\familydefault}{\mddefault}{\updefault}{$A/\lambda$}%
}}}}
\end{picture}%
}\\
Fig.~2.~Spatial degrees of freedom between two clusters\\ of area $A$ separated by distance $d$.
\end{center}
\vspace{5mm}

We see that the lower bound \eqref{eq:lb_dof} matches the upper bound \eqref{eq:ub_dof} in the case where the inter-cluster distance is smaller than or equal to the cluster radius ($d \le \sqrt{A}$), but nothing similar holds for $d \ge \sqrt{A}$. Our aim in the present paper is to close this gap and to show that in the regime where $d \ge \sqrt{A}$, the actual spatial degrees of freedom of the MIMO transmission do not exceed those found in \eqref{eq:lb_dof} (up to logarithmic factors). As a corollary, this would imply that when $d \ge A$, the number of degrees of freedom is bounded by 1.

In order to show this, we rely on an approximation whose validity is not fully proven here; it is however discussed in detail at the end of the paper. Our approach leads to an interesting result on the asymptotic behavior of the spectrum of random matrices that appear not to have been previously studied in the mathematical literature.

\section{Spatial degrees of freedom} \label{sec:spat}
Let us consider two square clusters of area $A$ separated by a distance $d$, one containing $n$ transmitters and the other containing $n$ receivers uniformly distributed in their respective clusters, as illustrated on Fig.~1. We are interested in estimating the number of spatial degrees of freedom of a MIMO transmission between the two clusters:
$$
Y_j = \sum_k \sqrt{F} \, h_{jk} \, X_k + Z_j, \quad j=1,\ldots,n,
$$
where $F$ is Friis' constant, the coefficients $h_{jk}$ are given by the line-of-sight fading model \eqref{eq:los_model} and $Z_j$ represents additive white Gaussian noise at receiver $j$. The distance $r_{jk}$ between node $j$ at the receiver side and node $k$ at the transmitter side is given by
\begin{equation} \label{eq:dist}
r_{jk} = \sqrt{(d + \sqrt{A} \, (x_j + w_k))^2 + A \, (y_j - z_k)^2}
\end{equation}
where $x_j,w_k,y_j,z_k \in [0,1]$ are normalized horizontal and vertical coordinates, as illustrated on Fig.~3.\\

\begin{center}
\scalebox{1}{\begin{picture}(0,0)%
\includegraphics{coord.pstex}%
\end{picture}%
\setlength{\unitlength}{1381sp}%
\begingroup\makeatletter\ifx\SetFigFont\undefined%
\gdef\SetFigFont#1#2#3#4#5{%
  \reset@font\fontsize{#1}{#2pt}%
  \fontfamily{#3}\fontseries{#4}\fontshape{#5}%
  \selectfont}%
\fi\endgroup%
\begin{picture}(9549,3562)(1639,-7211)
\put(4501,-7111){\makebox(0,0)[lb]{\smash{{\SetFigFont{6}{7.2}{\familydefault}{\mddefault}{\updefault}{$0$}
}}}}
\put(6301,-7111){\makebox(0,0)[lb]{\smash{{\SetFigFont{6}{7.2}{\familydefault}{\mddefault}{\updefault}{$d$}
}}}}
\put(9901,-6286){\makebox(0,0)[lb]{\smash{{\SetFigFont{6}{7.2}{\familydefault}{\mddefault}{\updefault}{$\sqrt{A} y_j$}
}}}}
\put(8701,-7111){\makebox(0,0)[lb]{\smash{{\SetFigFont{6}{7.2}{\familydefault}{\mddefault}{\updefault}{$\sqrt{A} x_j$}
}}}}
\put(2526,-6061){\makebox(0,0)[lb]{\smash{{\SetFigFont{6}{7.2}{\familydefault}{\mddefault}{\updefault}{$\sqrt{A} z_k$}
}}}}
\put(3151,-7111){\makebox(0,0)[lb]{\smash{{\SetFigFont{6}{7.2}{\familydefault}{\mddefault}{\updefault}{$-\sqrt{A} w_k$}
}}}}
\end{picture}
}\\
Fig.~3.~Coordinate system.
\end{center}

Assuming full channel state information and perfect cooperation of the nodes on both sides, the maximum number of bits per second and per Hertz that can be transferred reliably from the transmit cluster to the receive cluster over this MIMO channel is given by the following expression:
$$
C_n = \max_{Q \ge 0 \, : \, Q_{kk} \le P, \; \forall k} \log \det (I + H Q H^*),
$$
where $Q$ is the covariance matrix of the input signal vector $X=(X_1,\ldots,X_n)$ and $P$ is the power constraint at each node (in order to simplify notation, we choose units so that the other parameters, such as Friis' constant $F$, the noise power spectral density $N_0$ and the bandwidth $W$ do not appear explicitly in the above capacity expression).

In the sequel, we make the following {\em two assumptions}:\\

1) $d,A$ both increase\footnote{By ``increasing with $n$'', we mean that $A=n^\beta$ and $d=n^\gamma$ for some powers $\beta, \gamma>0$.} with $n$ and satisfy the relation $\sqrt{A} \le d \le A/\lambda$, which is the regime of interest to us (see Fig.~2).\\

2) $P=\frac{(d+\sqrt{A})^2}{n}$; because the average distance between two nodes in opposite clusters is $d+\sqrt{A}$ and because the MIMO power gain is of order $n$, this power constraint ensures that the SNR of the incoming signal at each receiving node is of order $1$ on average, so that the MIMO transmission operates at full power. Imposing this power constraint allows us to focus our attention on the spatial degrees of freedom of the system.\\

By choosing to transmit i.i.d.~signals (i.e.~taking $Q=PI$), we obtain
$$
C_n \ge \log \det (I + P \, HH^*)
$$
and using Paley-Zygmund's inequality, the following result was further shown in \cite{OLT13}.

\begin{thm} \label{thm:lb}
Under assumptions 1) and 2), there exists a constant $K_1>0$ such that
$$
C_n \ge \log \det (I + P HH^*) \ge K_1 \, \min \left( n, \frac{A/\lambda d}{\log (A/\lambda d)} \right)
$$
with high probability as $n$ gets large.
\end{thm}

This result shows that the number of spatial degrees of freedom of the MIMO transmission can reach $A/\lambda d$ (up to a logarithmic factor), when the number of nodes participating to the MIMO transmission is large enough.

As mentioned in the introduction, a natural question is whether it is possible to find a corresponding matching upper bound on the capacity. In order to answer this question, let us first observe that any matrix $Q$ satisfying the above constraints also satisfies $Q \le nPI$. Thus,
\begin{equation} \label{eq:ub_cap}
C_n \le \log \det(I + nP HH^*) = \sum_{k=1}^n \log(1 + \lambda_k)
\end{equation}
where $\lambda_1 \ge \lambda_2 \ge \ldots \ge \lambda_n$ are the eigenvalues of $nP HH^*$. The number of significant eigenvalues of $nP HH^*$ therefore determines the number of spatial degrees of freedom. The direct analysis of these eigenvalues appears to be difficult, so we proceed by approximating the matrix $nP HH^*$ by another matrix $GG^*$, easier to analyze.

\begin{claim} \label{claim:sim}
Let $m=A/\lambda d$ and $G$ be the matrix whose entries are given by
\begin{equation} \label{eq:approx}
g_{jk}= e^{-2\pi i m y_j z_k},
\end{equation}
where $y_j$, $z_k$, $1 \le j,k \le n$ are the same random variables as in expression \eqref{eq:dist}. Then under assumptions 1) and 2), the following approximation holds:
$$
\log \det(I + nP HH^*) = \log \det(I+GG^*) \, (1+o(1))
$$
with high probability as $n$ gets large.
\end{claim}

We discuss this approximation in detail in Section \ref{sec:discuss}. For the time being, observe first that by expression \eqref{eq:ub_cap}, the above approximation is equivalent to saying that the number of significant eigenvalues of $nP \, HH^*$ and $GG^*$ do not differ in order as $n$ gets large. Some numerical evidence of this fact is provided on Fig.~4 for a given set of parameters (a similar behavior is observed for a wide range of parameters).

\begin{center}
\includegraphics[width=6cm]{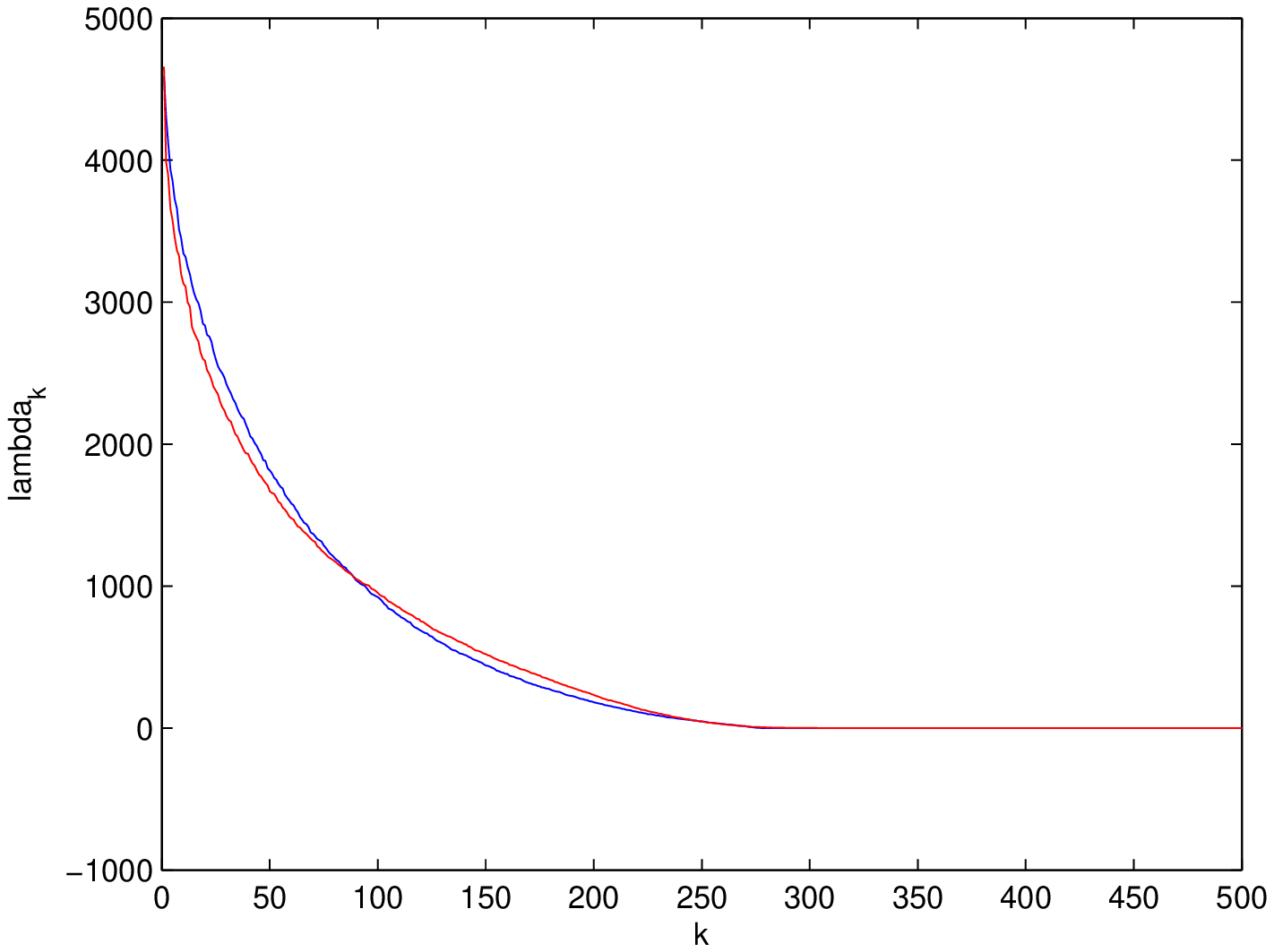}\\
Fig.~4. Eigenvalues of $nP HH^*$ (blue) and $GG^*$ (red) for the parameters $n=500$, $A=10'000m^2$, $d=300m$, $\lambda=0.1m$
(so $m=A/\lambda d \simeq 333$).
\end{center}

It can be observed on the figure that the eigenvalues drop to zero after a threshold of order $m=A/\lambda d$ for both matrices $nP HH^*$ and $GG^*$. Notice that because of assumption 1), we have $m=n^\delta$, where the power $\delta$ is a number greater than $0$.

The rest of the present section is devoted to the proof of the following statement.

\begin{thm} \label{thm:ub}
Let $m = A/\lambda d$  be such that\footnote{i.e.~$m=n^\delta$, where $\delta>1/2$.} $m \gg \sqrt{n}$. Then under assumptions 1) and 2), there exists a constant $K_2>0$ such that
$$
\log \det (I + GG^*) \le K_2 \, \min(n, m) \, \log n
$$
with high probability as $n$ gets large.
\end{thm}

This result shows that the lower bound found in Theorem \ref{thm:lb} is tight (provided Claim \ref{claim:sim} holds true and $m \gg \sqrt{n}$), which is saying that the number of spatial degrees of freedom of a MIMO transmission between two clusters of area $A$ separated by distance $d$ is of order $m=A/\lambda d$, up to logarithmic factors.

Let us also mention that applying the same technique as in \cite{OLT13}, the same matching lower bound on $\log \det (I + GG^*)$ can be found. This result on matrices $G$ of the form \eqref{eq:approx} is interesting in itself, as these do not appear to have been studied before in the random matrix literature.

\begin{proof}
First, observe that applying the same method as in \cite[Lemma 2.2]{OLT13}, the following concentration result can be shown: for all $\varepsilon>0$, there exists some constant $K>0$ such that 
$$
|\log \det (I + GG^*) - \EE(\log \det(I+GG^*))| \le K \, n^{1/2+\varepsilon}
$$
with high probability as $n$ gets large. As $m \gg \sqrt{n}$ by assumption, what remains to be shown is that there exists a constant $K_2>0$ such that
$$
\esp( \log \det (I + GG^*)) \le K_2 \, \min(n, m) \, \log n
$$
as $n$ gets large. 
Observe that this was the only part of the proof that requires $m \gg \sqrt{n}$. It follows that a sharper concentration bound would immediately yield a stronger result in Theorem \ref{thm:ub}.

In order to upperbound $\esp(\log\det(I+GG^*))$, let us now expand the determinant:
\begin{align*}
\lefteqn{\esp(\log\det(I + GG^*))}\\ 
& = \esp \bigg( \log \bigg( 1 + \sum_{k=1}^n \sum_{\mathcal{J} \subset \{1,\ldots,n\} \atop\abs{\mathcal{J}}=k} \det(G_{{\mathcal J} \times n} G_{{\mathcal J} \times n}^*) \bigg) \bigg)\\
& \le \log \bigg( 1 + \sum_{k=1}^n \sum_{\mathcal{J} \subset \{1,\ldots,n\} \atop\abs{\mathcal{J}}=k} \esp(\det(G_{{\mathcal J} \times n} G_{{\mathcal J} \times n}^*)) \bigg)
\end{align*}
where we used Jensen's inequality. Using the fact that the $y_j$ are i.i.d., we further obtain that $\esp(\det(G_{{\mathcal J} \times n} G_{{\mathcal J} \times n}^*))$ only depends on the size $k$ of the subset $\mathcal {J}$, so
\begin{align*}
\lefteqn{\esp(\log\det(I+GG^*))}\\
& \le \log \bigg( 1 + \sum_{k=1}^n {n \choose k} \, \esp(\det G_{k\times n} G_{k\times n}^*) \bigg)\\
& = \log \bigg( 1 + \sum_{k=1}^n {n \choose k} \esp \bigg( \sum_{\mathcal{I} \subset \{1,\ldots,n\} \atop \abs{\mathcal{I}}=k} \det(G_{k\times \mathcal{I}} G_{k\times \mathcal{I}}^*) \bigg) \bigg)\\
& = \log \bigg( 1 + \sum_{k=1}^n {n \choose k}^2 \esp(\det(G_{k\times k} G_{k\times k}^*)) \bigg)
\end{align*}
where we have used this time the Cauchy-Binet formula together with the fact that the $z_k$ are i.i.d. We thus see that in order to upperbound $\EE(\log \det(I+GG^*))$, it is enough to control $\esp(\det(G_{k \times k} G_{k \times k}^*))$, where $G_{k \times k}$ is the upper left $k\times k$ submatrix of $G$.

We will show that, similarly to what has been observed numerically for the eigenvalues $\lambda_k$, $\esp(\det(G_{k \times k} G_{k \times k}^*))$ drops rapidly for $k$ greater than a given threshold of order $m$, which will imply the result.

Using the definition of the determinant, we obtain
\begin{align*}
\lefteqn{\esp(\det(G_{k \times k} G_{k \times k}^*))}\\
& = \sum_{\sigma, \tau \in S_k} (-1)^{\abs{\sigma}+\abs{\tau}} \, \esp \bigg( \prod_{j=1}^{k} g_{j,\sigma(j)} \, \overline{g_{j,\tau(j)}} \bigg)\\
& = k! \sum_{\sigma \in S_k} (-1)^{\abs{\sigma}} \, \esp \bigg( \prod_{j=1}^k g_{jj} \, \overline{g_{j,\sigma(j)}} \bigg),
\end{align*}
which in turn leads to
\begin{align*}
\lefteqn{\esp(\det(G_{k \times k} G_{k \times k}^*))}\\
& = k! \sum_{\sigma \in S_k} (-1)^{\abs{\sigma}} \, \esp_Z \bigg( \prod_{j=1}^{k} \esp_Y (e^{-2 \pi i y_j (z_j -z_{\sigma(j)})}) \bigg)\\
& = k! \sum_{\sigma \in S_k} (-1)^{\abs{\sigma}} \, \esp_Z \bigg( \prod_{j=1}^{k} \frac{1-e^{-2 \pi i m (z_j-z_{\sigma(j)})}}{2 \pi i m (z_j-z_{\sigma(j)})} \bigg)\\
& = k! \, \EE_Z \bigg( \det \bigg( \left\{ \frac{1-e^{-2 \pi i m (z_j-z_l)}}{2 \pi i m (z_j-z_l)} \right\}_{1\le j,l\le k} \bigg) \bigg).
\end{align*}
Multiplying row $j$ by $e^{\pi i m z_j}$ and column $l$ by $e^{-\pi i m z_l}$, we reduce the problem to computing the following determinant:
$$
\esp_Z \left( \det \left( \left\{ \frac{\sin(\pi m (z_j-z_l))}{\pi m (z_j-z_l)} \right\}_{1\le j,l\le k} \right) \right).
$$
{\em Operators and Fredholm Theory.} The key observation is that the above expected value of the determinant can be seen as a classically studied quantity in the Fredholm theory of integral operators. This allows us to deduce precise estimates. A reference for the material discussed below is \cite{pinkus}.

Consider the continuous function $K_m (x,y)=\frac{\sin ( m (x-y))}{\pi(x-y)}$ on $[0,1]^2$ and the associated operator $K_m: C([0,1])\to C([0,1])$ defined as
$$
K_m \, \phi(x) = \int_0^1 \frac{\sin (m(x-y))}{\pi(x-y)} \, \phi(y) \, dy.
$$
The $p^{th}$ iterated kernel $K^{p}$ of an operator K is defined as $K^{1}=K$ and 
$$
K^{p}(x,y) = \int_{0}^{1} K^{p-1}(x,z) \, K(z,y) \, dz
$$
Associated to this is the $p^{th}$ trace of $K$:
$$
A_p = \int_{0}^{1} K^{p}(x,x) \, dx
$$
Define as well the compound kernel $K_{[p]}\in C([0,1]^{2p})$ as
$$
K_{[p]}({\bf x},{\bf y}) = \det \begin{pmatrix} K(x_1,y_1) & K(x_1,y_2) & \ldots & K(x_1, y_p)\\
\vdots & \vdots & \ddots &\vdots\\
K(x_p,y_1) & K(x_p,y_2) & \ldots & K(x_p,y_p) \end{pmatrix}
$$
for ${\bf x}=(x_1,\ldots, x_p)$ and ${\bf y}=(y_1,\ldots, y_p)$. In this notation, the quantity we are interested in is
$$
\esp_Z \left( \det \left( \left\{ \frac{\sin(\pi m (z_j-z_l))}{\pi m (z_j-z_l)} \right\}_{1\le j,l\le k} \right) \right) = k! \, \frac{1}{m^k} \, d_k,
$$
where
$$
d_k = \frac{1}{k!} \int_{[0,1]^k} K_{m,[k]}({\bf x},{\bf x}) \, dx_1 \cdots dx_k.
$$
Since $K_m$ is a compact operator, it has a discrete spectrum $\mu_1 \ge \mu_2 \ge \ldots$ It is immediate to see that $A_{p}=\sum \mu_{i}^{p}$. Furthermore, the quantities $d_k$ and $A_p$ are related by the following recurrence relation, which follows from expanding the determinant in the definition of $K_{[p]}$ and regrouping equal terms (see \cite{pinkus}).
$$
k \, d_k = \sum_{p=1}^k (-1)^{p-1} \, A_p \, d_{k-p}.
$$
Using the fact that $A_{p}=\sum \mu_{i}^{p}$, it can be seen that the only solution to the above recurrence (with $d_{0}=1$) is 
\begin{equation} \label{eq:product}
d_k = \sum_{i_1<i_2<\ldots<i_k} \mu_{i_1} \, \mu_{i_2} \cdots \mu_{i_k}. 
\end{equation}
We conclude that it is sufficient to estimate the eigenvalues of the operator $K_m$ in order to upperbound $d_k$.

Since the kernel $K_m$ is translation invariant, it can be defined equivalently on $[-1/2,1/2]$, and this new operator has the same eigenvalues. This operator is called the sinc kernel and is well known in signal processing, since it is the Fourier transform of the indicator function. It was originally studied by D.~Slepian in \cite{slepian} and precise estimates exist on the behavior of its eigenvalues. In particular, we have the following recent result from \cite[Theorem 3]{pswf}:

\begin{thm}
Let $\delta>0$. There exists $M \ge 1$ and $c>0$ such that, for all $m\ge 0$ and $k \ge \max(M,cm)$, 
$$
\mu_k \le e^{-\delta \, (k-c m)}.
$$
\end{thm}

This theorem essentially says that the eigenvalues $\mu_k$ decay exponentially for $k \ge cm$. The direct consequence of this is that
$d_k$ decays like $\exp(-\delta \, (k-cm)^2/2)$ for $k\ge cm$, as we show below.

Indeed, it follows that if we take $k$ sufficiently large (i.e.~larger than $cm$), the sum in \eqref{eq:product} always contains at least one term with exponential decay. Recalling that all eigenvalues $\mu_k$ are bounded by $1$, we obtain
\begin{align*}
d_k & =\sum_{i_1<\ldots <i_k} \mu_{i_1} \cdots \mu_{i_k} \\
& \le \sum_{i_1<\ldots <i_{k-1}} \min(e^{-\delta(i_1-c m)},1) \cdots \min(e^{-\delta(i_{k-1}-c m)},1)\\
& \hspace{1cm} \times \sum_{i_k=k}^{\infty} e^{-\delta(i_k-c m)},
\end{align*}
as $i_k$ is necessarily greater than or equal to $k$ in the above sum. Now, define $K=k-c m$ and observe that
$$
\sum_{i_k=k}^{\infty} e^{-\delta(i_k-c m)} = \frac{e^{-\delta (k-c m)}}{1-e^{-\delta}} \le C \, e^{-\delta K},
$$
for some constant $C>0$. So by induction, we further obtain
$$
d_k \le C^K \prod_{j=1}^K e^{-\delta j}  \le C^K \, e^{- \delta K^2/2}.
$$
This gives us the estimate we were after for $d_k$:
$$
d_k \le \begin{cases} 1, & \text{if } k \le c m,\\ C^k \, e^{-\delta \, (k-c m)^2/2}, &\text{if } k > c m. \end{cases}
$$
Gathering all estimates together, we finally obtain
\begin{align*}
\lefteqn{\esp(\log\det(I+GG^*))}\\
& \le \log \bigg( 1+ \sum_{k=1}^n {n \choose k}^2 \, \frac{k!^2}{m^k} \, d_k \bigg) \le \log \bigg( 1+ \sum_{k=1}^n n^{2k} \, d_k \bigg)\\
& \le \log \bigg( 1+ \sum_{k=1}^{c m} n^{2k} + \sum_{k=c m+1}^n n^{2k} \, C^k \, e^{-\delta \, (k-c m)^2/2} \bigg)\\
& \le (c m +1) \, \log n + O(1),
\end{align*}
which concludes the proof of Theorem \ref{thm:ub}.
\end{proof}

\section{Discussion} \label{sec:discuss}
Our aim in the following is to provide a justification for Claim \ref{claim:sim}. Let us first recall the definition of both $H$ and $G$:
$$
h_{jk} = \frac{e^{2 \pi i r_{jk} / \lambda}}{r_{jk}} \quad \text{and} \quad g_{jk}= e^{- 2\pi i m y_j z_k},
$$
where $m=A/\lambda d$ and
$$
r_{jk} = \sqrt{(d + \sqrt{A} \, (x_j + w_k))^2 + A \, (y_j - z_k)^2}.
$$
Given the chosen power constraint $P$ and the fact that $d \ge \sqrt{A}$, it follows that the amplitude of the normalized fading coefficent $\sqrt{nP} \, h_{jk}$ is of order $1$, matching that of $g_{jk}$. Let us now compare the phases of these two coefficients. Using a Taylor approximation to quadratic order around $0$ in the variable $(y_j-z_k)$, we get
\begin{align*}
\lefteqn{r_{jk} \simeq d + \sqrt{A} \, (x_j + w_k) + (A/d) \, (y_j-z_k)^2/2}\\
& = d + \sqrt{A} \, (x_j+w_k) + (A/d) \, (y_j^2/2 + z_k^2/2) - (A/d) \, y_j z_k.
\end{align*}
Hence,
$$
e^{2 \pi i r_{jk} /\lambda} \simeq \widetilde{h}_{jk} := e^{2 \pi i (u_j + v_k - (A/\lambda d) \, y_j z_k)},
$$
where
\begin{align*}
\begin{cases} u_j = (d/2 + \sqrt{A} \, x_j + (A/d) \, y_j^2/2)/\lambda,\\
v_k = (d/2 + \sqrt{A} \, w_k + (A/d) \, z_k^2/2)/\lambda. \end{cases}
\end{align*}
Notice moreover that the eigenvalues of $\widetilde{H} \widetilde{H}^*$ do not depend on the particular values of the $u_j$'s or $v_k$'s; they are therefore the same as the eigenvalues of $GG^*$, which shows finally that
$$
\sqrt{nP} \, h_{jk} \simeq g_{jk}, \quad \forall j,k.
$$
This entry-by-entry approximation adds some plausibility to Claim \ref{claim:sim}.

\section{Conclusion and perspectives} \label{sec:conc}
The goal of this work is to give precise estimates on the number of spatial degrees of freedom in large MIMO systems in a line-of-sight environment. An upper bound for a model closely related to the line-of-sight model has been given, and the similarity of the models is supported numerically. As such, it remains to be shown that the eigenvalues of the two models are indeed very close, in order to bound $|\log \det (I + nP \, HH^*) - \log \det(I+ GG^*)|$; this is work in progress.

As a by-product, the spectral properties of random matrices $G$ of the form $g_{jk}=e^{-2 \pi i m y_j z_k}$ have been studied in this paper. These matrices are not unrelated to both Vandermonde matrices and random DFT matrices that appear in other contexts in the literature on wireless communications \cite{NCV08,RD09,TW11,TCSV10} and compressed sensing \cite{candes,farrell}, respectively:

- Vandermonde matrices are simply obtained by choosing $y_j=j/n$ deterministic instead of uniform and i.i.d.~on $[0,1]$. In this case, the matrix entries are given by $e^{-2 \pi i m j z_k / n}$. Our analysis technique does not allow us to reach the same conclusions directly for these matrices, but we believe this method to be fruitful for further developments.  

- Random DFT matrices are obtained from the above Vandermonde matrices by further replacing $m z_1, \ldots, m z_n$ by a subset $\{1_1,\ldots,l_m\}$ of $m$ integers chosen uniformly at random in $\{1,\ldots,n\}$, so the matrix entries become $e^{-2 \pi i j l_k/n}$. Again, our technique can be applied in this setup.

\bibliographystyle{plain}
\bibliography{dof_los_mimo.bib}

\end{document}